\documentclass[aps,prl,twocolumn,showpacs]{revtex4}
\usepackage{graphicx}
\begin{document}
\title{ Universal mean moment rate profiles of earthquake ruptures }
\author{ Amit P. Mehta }
\email[]{mehtafree@yahoo.com}
\affiliation{ Department of Physics, University of Illinois at
Urbana-Champaign, 1110 West Green Street, Urbana, IL 61801-3080 }
\author{Karin A. Dahmen}
\email[]{dahmen@uiuc.edu}
\affiliation{ Department of Physics, University of Illinois at
Urbana-Champaign, 1110 West Green Street, Urbana, IL 61801-3080 }
\author{Yehuda Ben-Zion}
\email[]{benzion@terra.usc.edu}
\affiliation{ Department of Earth Sciences, University of Southern CA,
Los Angeles, CA 90089-0740 } 


\newcommand{\be}{\begin{equation}}
\newcommand{\ee}{\end{equation}}
\newcommand{\bea}{\begin{eqnarray}}
\newcommand{\eea}{\end{eqnarray}}
\newcommand{\la}{\langle}
\newcommand{\ra}{\rangle}

\begin{abstract}
Earthquake phenomenology exhibits a number of power law distributions
including the Gutenberg-Richter frequency-size statistics and the
Omori law for aftershock decay rates. In search for a basic model that
renders correct predictions on long spatio-temporal scales, we discuss
results associated with a heterogeneous fault with long range
stress-transfer interactions.  To better understand earthquake
dynamics we focus on faults with Gutenberg-Richter like earthquake
statistics and develop two universal scaling functions as a stronger
test of the theory against observations than mere scaling exponents
that have large error bars. Universal shape profiles contain crucial
information on the underlying dynamics in a variety of systems. As in
magnetic systems, we find that our analysis for earthquakes provides a
good overall agreement between theory and observations, but with a
potential discrepancy in one particular universal scaling function for
moment-rates. The results point to the existence of deep connections
between the physics of avalanches in different systems.
\end{abstract}


\maketitle

\section{ Introduction } 

Earthquake phenomenology is characterized by several power law
distributions. The most famous of these are the frequency-size
distributions (i.e. histograms) of regional and global
earthquakes \cite{GR,utsu2}, and the modified Omori law for the aftershock
decay rate around large rupture zones \cite{utsu,kiss}.  Using the
seismic moment $M_0$ for the earthquake size, the frequency-size
distributions (or moment histogram) has the form (e.g.,
\cite{benzion})
\begin{equation}
n(M_0) \sim M_0^{-1-\beta}
\label{eq1}
\end{equation}
where $M_0\sim \sum_i\Delta u_i\Delta A_i$ with $\Delta u_i$ and
$\Delta A_i$ being the local slip and rupture area during an
earthquake, respectively. A related, more commonly used form in terms
of earthquake magnitude $M$ is
\begin{equation}
log(n(M)) = a-bM
\label{eq2}
\end{equation}
where $n(M_0)dM_0=n(M)dM$, the constant $a$ characterizes the overall
rate of activity in a region, and the "$b$-value" gives the relative
rates of events in different magnitude ranges. Using the observed
moment-magnitude scaling relation $M \sim 2/3log(M_0)$ for large
earthquakes \cite{benzion,hanks,utsu2}, the exponent $\beta$ of (\ref{eq1}) is
related to the $b$-value of (\ref{eq2}) as $b=1.5\beta$.
The modified Omori law for aftershock
decay rates is:
\begin{equation}
\Delta N/\Delta t \sim K/(t+c)^p
\label{eq3} 
\end{equation}
where $N$ is the cumulative number of aftershocks, $t$ is the time
after the mainshock, and $K$, $c$, and $p$ are empirical constants.
The exponents in (\ref{eq1}) and (\ref{eq3}) are stable for data
collected over large space-time domains, with some clear deviations
from global averages related to faulting type and regional properties
\cite{utsu2}.  For example, the $b$-values of strike-slip, thrust, and
normal earthquakes with depth $\le50$ km in the global Harvard
catalogue are about 0.75, 0.85, and 1.05, respectively \cite{frolich}.
As another example, regions with high heat flow often have short
aftershock sequences with relatively large exponent (e.g., $p >
1.25$), while regions with low heat flow have long aftershock
sequences with low exponent (e.g., $p < 0.9$) \cite{utsu,kiss}.  The
association of earthquake statistics with power law relations like Eq.
(\ref{eq1}) and Eq. (\ref{eq3}) led some to suggest that earthquake
dynamics is associated with an underlying critical point
\cite{zapperi,fisher,bak,sornet,Klein}.  
However, power
law distributions can be generated by many other mechanisms
\cite{schroeder,sornette} and it is important to develop criteria that
can provide stronger evidence for or against the association of
earthquakes with criticality. 

Recently, enough data have been collected to extract statistics of
earthquakes on individual fault zones occupying long (order 100 km)
and narrow (order 10 km) regions of space.  Wesnousky and
collaborators \cite{wes,stirling} found that the frequency-size
statistics of earthquakes on highly irregular fault zones, with many
offsets and branches, as the San Jacinto fault in California, are also
described by the Gutenberg-Richter power law relation up to the
largest events.  However, relatively regular fault zones (presumably
generated progressively with increasing accumulated slip over time),
such as the San Andreas fault in California, display power-law
frequency-size statistics only for small events. These occur in the
time intervals between roughly quasi-periodic earthquakes of a much
larger ``characteristic'' size that is related to large-scale
dimensions of the fault zone \cite{benzion,wes,benzion2,benzion4}.
(If the ratio of the mean divided by the standard deviation of the
distribution of time intervals between characteristic earthquakes is
larger than 1, the distribution is referred to as quasi-periodic
\cite{benzion4}).  Earthquakes of intermediate magnitude are typically
not observed on these faults (other than, perhaps, during aftershock
sequences). The corresponding frequency size statistics are called the
"characteristic earthquake" distribution \cite{benzion,wes}.

Previously these two types of behavior on individual fault zones have
been modeled as statistics close-to and far-from an underlying
critical point \cite{fisher,dahmen}, using a model for a strike-slip
fault that incorporates long-range interactions and strong
heterogeneities \cite{benzion2,benzion4}.  The different dynamic
regimes were associated with a competition between failure-promoting
effects of elastic stress-transfer or dynamic weakening, and the
opposing effect of strength inhomogeneities in the fault
structure. Fisher et al. \cite{fisher} found that near the critical
point the frequency-size statistics follow a power law distribution
(with a cutoff at large magnitude), with the same scaling exponent of
observed data for strike-slip faults (i.e., a $b$-value of 0.75). A
similar form of frequency-size statistics and predicted $b$-value were
obtained also for a critical parameter value in a stochastic branching
model \cite{vere}.

To provide an improved understanding of earthquake dynamics that can
suggest additional observables, we focus on faults with
Gutenberg-Richter like earthquake statistics (i.e. near-critical
behavior) and develop two universal scaling {\em functions} associated
with mean moment-rate time profiles at either fixed total moment or
fixed total earthquake duration. Universal scaling functions (or shape
profiles) give important information on the underlying dynamics, and
may be found in solar flares in astrophysics \cite{lu}, price
fluctuations in financial markets \cite{bouchaud}, Barkhausen noise in
magnets \cite{mehta}, and, as shown here, also in earthquake
phenomenology. If the behavior of fault zones with earthquakes
following the Gutenberg-Richter statistics is indeed critical, then
the shapes of these functions should be as universal as the exponent
$\beta$ in Eq. (\ref{eq1}). Comparing the scaling functions in our 
earthquake model to observations constitutes a much stronger test 
of the theory than merely comparing a discrete, finite set of 
critical exponents.

In this study we compute the scaling functions for both model
predictions and observational data and compare the results.  In
section \ref{model} of the paper we review the model. In sections
\ref{dynam} to \ref{phasedia}, we extend the model, while also
introducing an extended phase diagram for the model dynamics, and the scaling
behavior on long length scales. In section \ref{profile} we introduce
the universal scaling functions and their scaling forms, and in
section \ref{exposcale} we extract the functions from both simulation
and observational data. Finally, in section \ref{conclude} we discuss
the results and emerging new questions.

\section{\label{model} Earthquake Model} 

The model we use was developed originally by Ben-Zion and Rice
\cite{benzion2,benzion4}, who suggested that a narrow irregular
strike-slip fault zone of horizontal length $L$ and vertical depth $W$
may be represented by an array of $N \sim LW$ cells in a two
dimensional plane, with constitutive parameters that vary from cell to
cell to model the disorder (offsets etc.) of the fault zone structure
(FIG. \ref{planar}). The cells represent brittle patches on the
interface between two tectonic blocks that move with slow transverse
velocity $v$ in the $x$ direction at a great distance from the fault.
The interaction between cells during slip events is governed by 3-D
elasticity and falls off with a distance $r$ from the failure zone as
$\frac{1}{r^3}$. These interactions are sufficiently long range that
scaling in mean field theory (where the interaction range is set to
infinity) becomes exact, up to logarithmic corrections, in the
physical fault dimension ($d=2$) \cite{benzion2,benzion4,fisher}.  

In mean field theory, the local stress $\tau_i$ on a given cell $i$ is
\cite{benzion2}:
\begin{eqnarray}
\tau_i &=& J/N\sum_j(u_j-u_i)+K_L(vt-u_i)\\
&=& J\bar{u}+K_Lvt-(K_L+J)u_i
\label{eq4}
\end{eqnarray}
where $u_i$ is the total offset of cell in the horizontal $x$
direction, $\bar{u}=\sum_ju_j/N$ is the average displacement, $J/N$ is
the mean-field elastic coupling strength between cells, and $K_L \sim
1/\sqrt{N}$ is the loading stiffness \cite{dahmen} of the tectonic
blocks that move far away from the fault with relative velocity $v$.
Initially the stresses $\tau_i$ are randomly distributed with
$\tau_{a,i} \le \tau_i \le \tau_{s,i}$, where $\tau_{s,i}$ is a fixed
local {\em static} failure threshold stress and $\tau_{a,i}$ is the
fixed local arrest stress.  The distributions of static failure
stresses and arrest stresses represent the heterogeneity or
geometrical disorder in the fault system.  The differences between the
failure and arrest stresses give the local distribution of stress
drops during brittle failures; the earthquake dynamics depend only on
the stress drop distribution (no stochasticity). In addition, the
scaling behavior of the system is not sensitive to the exact form of
the distributions as long as they are bounded and
$\tau_{a,i}<\tau_{s,i}$. We choose a compact distribution for
$\tau_{a,i}$, such that $p(\tau_{a,i})=3(W^2-4\tau_{a,i}^2)/(2W^3)$
for $-W/2 \le \tau_{a,i} \le W/2$ and $0$ outside of these bounds.
Also, we look at the low disorder limit where $W \ll \tau_{s,i}$,
where we choose $\tau_{s,i} = 1$, so all cells will fail at this
point.

The fault is stuck while the stress on each cell is increased
uniformly as $d\tau_i/dt = K_Lv$ as a result of the external loading
which is increased adiabatically (that is, we take the limit $v
\rightarrow 0$). When the stress on a cell reaches its failure
threshold $\tau_{s,i}$, the cell slips by the amount:
\begin{equation}
\Delta u_i = (\tau_{s,i}-\tau_{a,i})/(K_L+J). 
\end{equation}
This stress drop is uniformly redistributed to all other cells 
(employed in the mean field approximation) by an amount:
\be
\delta \tau_j = (c/N)(\tau_{s,i}-\tau_{a,i}), j \ne i
\ee
where $c \equiv J/(K_L+J)$ is the conservation parameter which gives
the fraction of the stress drop of a cell that is retained in
the system after it slips \cite{dahmen}.  The resulting stress
increase on the other cells can cause some of them to slip as well,
leading to an avalanche of cell slips, or an earthquake.

\section{\label{dynam} Dynamical Weakening and Strengthening}

\begin{figure}
\includegraphics[angle= 270,scale= .75]{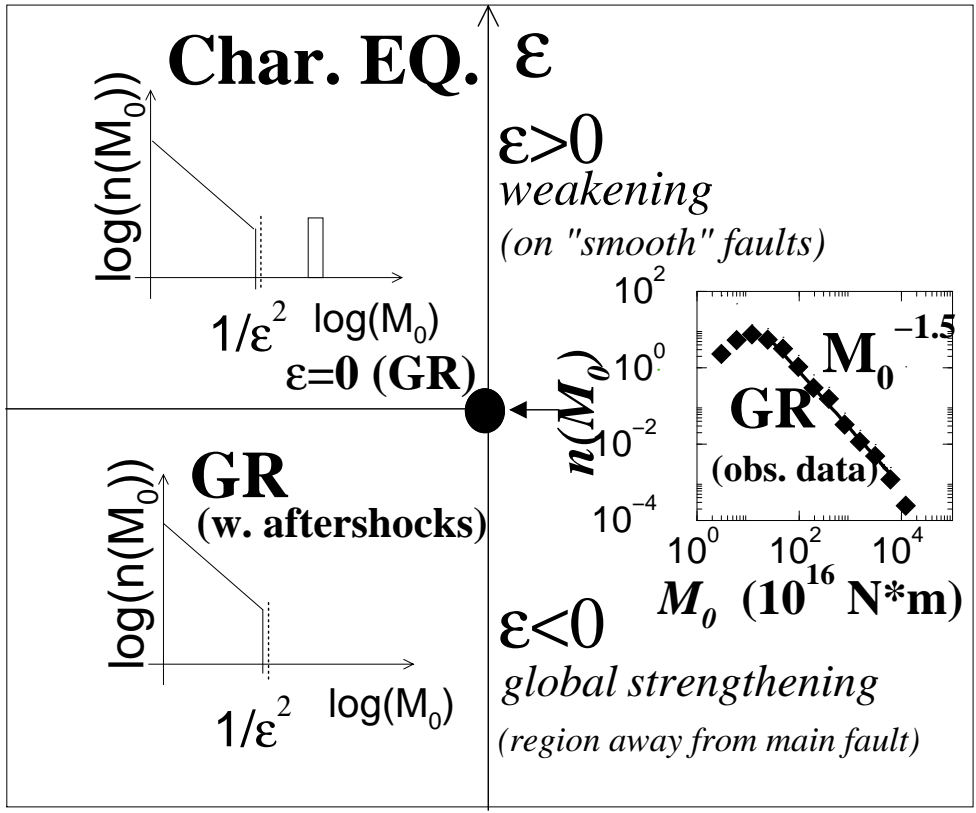}
\caption{\label{phase} Phase diagram of the model (see text and
\protect\cite{dahmen} for details).  }
\end{figure}

The model includes dynamic weakening effects during the failure
process \cite{benzion2,benzion4}: after an initial slip in an
earthquake, the strength of a failed cell is reduced to a {\em
dynamical} value:

\be
\tau_{d,i} \equiv \tau_{s,i}-\epsilon(\tau_{s,i}-\tau_{a,i}), 
\ee

with $0\le\epsilon\le1$ parameterizing the relative importance of
dynamical weakening effects in the system. This weakening represents
the transition from static friction to dynamic friction during the
rupture and the strength of a failed cell remains at the dynamic value
throughout the remainder of the earthquake. In the time intervals
between earthquakes all failure thresholds heal back to their static
value $\tau_{s,i}$. Fisher et al. \cite{fisher} found that at exactly
$\epsilon = 0$ the model produces a power law distribution of
earthquake moments $M_0$ following equation (\ref{eq1}), cutoff by the
finite fault size, with an analytical exponent $\beta = 1/2$ (FIG.
\ref{phase}). This corresponds to a $b$-value of 0.75, close to that
associated with observed earthquakes on strike-slip faults
\cite{frolich}. The power law scaling of the frequency-size statistics
and other variables \cite{fisher} indicates that the model with
$\epsilon = 0$ operates at a critical point.  In contrast, for a
finite weakening $\epsilon > 0$ the model produces the characteristic
earthquake distribution, with power law statistics for the small
events up to a cutoff moment that scales like:
\be
M_0^{cutoff}\sim 1/\epsilon^2, 
\ee
and quasi-periodically recurring large characteristic events that
scale with the fault size ($M_0 \sim (LW)^{3/2}$).

\begin{figure}
\scalebox{0.75}{
\includegraphics*[38,80][565,760]{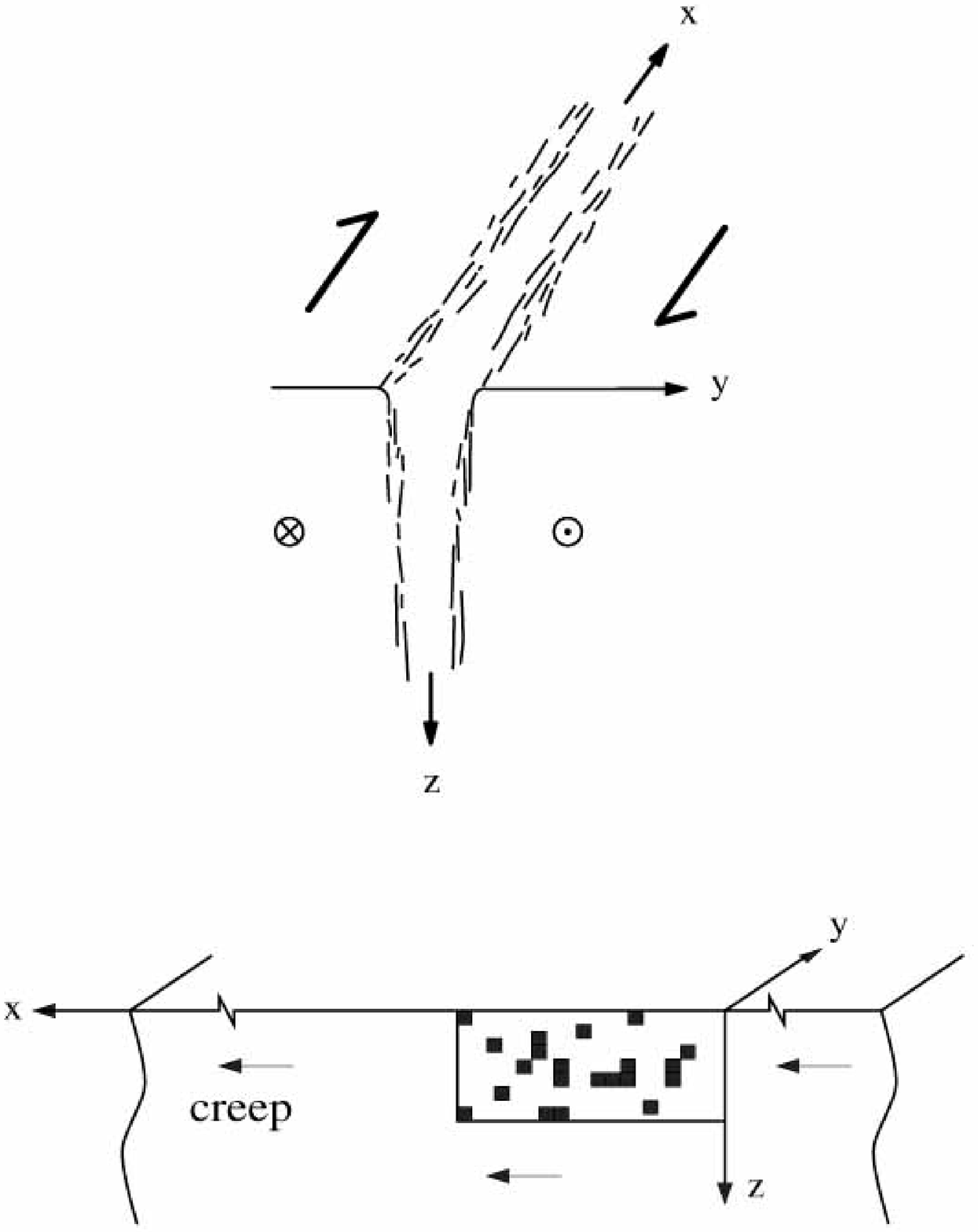}}
\caption{\label{planar} A planar representation of a 3-D segmented
fault zone by a 2-D heterogeneous fault embedded in a 3-D solid
\protect\cite{benzion2,benzion4}.}
\end{figure}

The model can be expanded further to include dynamic strengthening
represented by $\epsilon < 0$. Ben-Zion and Sammis \cite{benzion5}
summarized multidisciplinary observations which indicate that brittle
failure of rock has an initial transient phase associated with 
strengthening, distributed deformation, and creation of
new structures. Detailed frictional studies also show an initial 
strengthening phase associated with the creation of a new population
of asperity contacts \cite{benzion,dieterich}. In our model 
(FIG. \ref{planar}) we associate $\epsilon < 0$ with regions
off the main fault segments that are in an early deformation stage. To
capture basic aspects of brittle deformation on such regions in the
three-dimensional volume around the main fault (FIG. \ref{planar}), we
change the model as follows: when any cell $i$ slips during an
earthquake, and thereby reduces its stress by $\Delta\tau_i \equiv
\tau_{f,i}-\tau_{a,i}$, the failure stress $\tau_{f,j}$ of {\em every}
cell $j=1,...,N$ is {\em strengthened} by an amount
$|\epsilon|\Delta\tau_i/N$. Once the earthquake is complete, the
failure stress of each cell is slowly lowered back to its original
value. This represents in a simple way the brittle deformation that
occurs during an earthquake in the off-fault regions, which are first
in a strengthening regime, compared to the main fault,
and then have a weakening process. The
events that are triggered as the failure stresses are lowered in the
weakening period are referred to as {\em aftershocks}. The occurrence
of aftershocks in this version of the model for off-fault regions is
in agreement with the observation that a large fraction of observed
aftershocks typically occur in off fault regions \cite{liu}. For this
version of the model with $\epsilon < 0$, both the primary earthquakes
(i.e., main shocks) and the triggered aftershocks are distributed
according to the Gutenberg-Richter distribution, up to a cutoff moment
scaling as $1/\epsilon^2$. Assuming that the increased failure stress
thresholds $\tau_{f,i}$ are slowly lowered with time as $log(t)$
towards their earlier static values $\tau_{s,i}$, and that the
stresses are distributed over a wide range of values, we show analytically
in Appendix \ref{Omori} that the temporal decay of aftershock
rates at long times is proportional to $1/t$, as in the modified Omori
law of Eq. (\ref{eq3}) with $p=1$ \cite{utsu,kiss,benzion}.

Remarkably, the long length scale behavior of this model can be shown
to be the same as the behavior of the model given in Eq. (\ref{eq4})
with an added ``antiferroelastic'' term ($-|\epsilon|J\bar{u}$):
\begin{equation}
\tau_i = J\bar{u}+K_Lvt-(K_L+J)u_i-|\epsilon|J\bar{u} \,.
\label{eq5}
\end{equation}
In Eq. (\ref{eq5}) every time a cell fails, it slips by an amount
$\Delta u_i$ that leads to stress loading of the other cells, lessened
by $|\epsilon|J\Delta u_i/N$ compared to our original model (Eq.
(\ref{eq4})). On the other hand, in the global strengthening model
(described above) when a cell slips the failure stresses of all cells
are strengthened by $|\epsilon|J\Delta u_i/N$. On long length scales
the global strengthening of the failure stress has equivalent effects
on the earthquake statistics as the dissipation of the redistributed
stress, up to corrections of order $O(1/N)$, so the scaling behavior
for large events of both models are the same. Moreover, Eq.
(\ref{eq5}) can be rewritten as:
\begin{equation}
\tau_i = J[1-|\epsilon|][\bar{u}-u_i]+K_Lvt-[K_L+J|\epsilon|]u_i \, .
\label{eq5a}
\end{equation}
We can now absorb $|\epsilon|$ by defining $J^{\prime}=J(1-|\epsilon|)$ and
$K^{\prime}_L=K_L+J|\epsilon|$.  Rewriting Eq. (\ref{eq5a}) with the new
definitions, and dropping the $|\epsilon|$ contribution in
$[K^{\prime}_L-J|\epsilon|]vt$ since $v\rightarrow 0$, we find:
\begin{equation}
\tau_i = J^{\prime}\bar{u}+K^{\prime}_Lvt-(K^{\prime}_L+J^{\prime})u_i \, .
\label{eq5b}
\end{equation}
Therefore we recover Eq. (\ref{eq4}) with $J\rightarrow J^{\prime}$
and $K_L\rightarrow K^{\prime}_L$. This amounts to changing the stress
conservation parameter $c$ (from reference \cite{dahmen}). For Eq.
(\ref{eq5b}):
\begin{equation}
c=J^{\prime}/(K^{\prime}_L+J^{\prime}) = 1-|\epsilon|
\label{cparam}
\end{equation}
where $K_L\rightarrow 0$ since we are concerned with the adiabatic
limit.  We also know (from reference \cite{dahmen}) that the
cutoff $S_{cf}$ for the Gutenberg-Richter distribution scales as
$S_{cf}\sim 1/(1-c)^2$.  Thus, from Eq. (\ref{cparam}) we find that
the cutoff for Eq. (\ref{eq5}) will scale as $\sim1/|\epsilon|^2$.

\section{Mapping to Single Interface Magnet Model}

The mean field version of the single interface magnet model with
infinite range antiferromagnetic interactions is given by
\cite{zapperi2,durin}:
\begin{equation} 
\dot{h}_i(t) = J[\bar{h}-h_i(t)]+H(t)-k\bar{h}+\eta_i(h)
\label{mageqn}
\end{equation}
where $h_i(t)$ is the position of the domain wall, $H(t)$ is the
external driving field, $k$ is the coefficient of the
antiferromagnetic term, and $\eta_i(h)$ is the pinning field.  In
the paper by Fisher et al. \cite{fisher} it has been shown that the
scaling behavior on long length scales resulting from Eq. (\ref{eq5}),
without the $-|\epsilon|J\bar{u}$ term, is same as that of
Eq.  (\ref{mageqn}) without the antiferromagnetic term $-k\bar{h}$.
Furthermore, upon inspection we see the following correspondence
between the single interface magnet model (Eq. (\ref{mageqn})), and
the mean field earthquake model (Eq. (\ref{eq5})):

\be
-k\bar{h} \Longleftrightarrow -|\epsilon|J\bar{u}
\ee

In other words, the coefficient of the antiferromagnetic term $k$
plays the same role in the magnet model (Eq. (\ref{mageqn})), as the
coefficient of strengthening $|\epsilon|J$ does in the earthquake model
(Eq. (\ref{eq5})).

\section{\label{phasedia} Phase Diagram}

The regimes with various statistics produced by the model are
summarized by the phase diagram given in FIG. \ref{phase}. The range
$\epsilon > 0$ corresponds to ``mature'' localized faults with a
weakening rheology and characteristic earthquake statistics. The value
$\epsilon = 0$ corresponds to ``immature'' strongly inhomogeneous
fault zones with Gutenberg-Richter statistics. Finally, the range
$\epsilon < 0$ corresponds to the fracture network away from the main
fault, characterized by strengthening due to the creation of new
structures and associated emerging aftershocks.  It may be surprising
that the discussed simple model can capture many of the essential
general features of earthquake statistics (or other systems with
avalanches, such as driven magnetic domain walls). This can be
understood through the renormalization group \cite{Nature,binney}, a
powerful mathematical tool to coarse grain a system and extract its
effective behavior on long space-time scales. Many microscopic details
of a system are averaged out under coarse graining, and universal
aspects of the behavior on long scales depend only on a few basic
properties such as symmetries, dimensions, range of interactions,
weakening/strengthening, etc. When a model correctly captures those
basic features, the results provide proper predictions for statistics,
critical exponents, and universal scaling functions near the critical
point. Consequently, many models that are in the same universality
class lead to the same statistics and exponents
\cite{Nature,fisher,dahmen,binney}.  The universal scaling functions
around the critical point, discussed in the next section, provide
additional information that can be used to distinguish between
different models and universality classes.

\section{\label{profile} Moment Rate Shapes }

In this section we focus on fault zones with Gutenberg-Richter power
law statistics, modeled by systems at or close to the $\epsilon = 0$
critical point. Recent analysis allowed researchers to obtain the
moment rate $dm_0(t)/dt$, which gives the slip on a fault per unit
time during the propagation of earthquake rupture, for hundreds of
large seismic events recorded on global networks \cite{bilek,Houston}.
The moment rates are derived from inversions of teleseismically recorded
seismograms on a global seismic network \cite{Ruff}.
Motivated by works on statistical physics of magnetic systems 
discussed in the previous chapter (see also \cite{Nature,mehta}), we are
interested in studying the event-averaged moment rate time profile
(FIG. \ref{Mplot}) for earthquakes with given total moment $M_0$,
denoted with $\la dm_0(t|M_0)/dt \ra$ and 
the event-averaged moment rate time profile
(FIG.  \ref{Tplot}) for earthquakes with given duration $T$, denoted
with $\la dm_0(t|T)/dt \ra$.
Here $m_0(t|T)$ is the (cumulative)
moment at time $t$ of the propagating earthquake of total duration
$T$, and $m_0(t|M_0)$ is the cumulative moment at time $t$ of the
earthquake of total moment $M_0$. Theoretical analysis of phase
diagrams similar to that shown in FIG. \ref{phase} implies that near
the critical point there should be, in addition to scaling exponents,
also universal scaling functions (up to a rescaling of the ordinate
and abscissa) \cite{mehta}.  In our model the two scalable functions
of interest, $\la dm_0(t|M_0)/dt \ra$ and $\la dm_0(t|T)/dt \ra$, obey
respectively the following scaling relations \cite{Matt1,fisher2}:

\begin{equation}
\la dm_0(t|M_0)/dt \ra/M_0^{1/2}\sim f(t/M_0^{1/2})
\label{eq6}
\end{equation}

and 

\begin{equation}
\la dm_0(t|T)/dt \ra\sim g(t/T)
\label{eq7}
\end{equation}

We determined these scaling functions from corresponding results for
magnets, using the fact that our mean field version of the Ben-Zion
and Rice model of Eq. (\ref{eq4}) \cite{fisher,dahmen} is in the same
universality class (i.e. has the same universal behavior on long
length scales) as the above mentioned model for domain wall motion in
magnets. 

\section{\label{exposcale} Exponents and Data Collapses }

We compare the observation results with our model and find remarkable
agreement in most cases. The frequency-moment distribution, $D(M_0)
\sim M_0^{-1-\beta}$ of the observed data \cite{bilek} has (inset of
FIG. \ref{phase}) three decades of scaling and an exponent of
$\beta=1/2\pm0.05$, in close agreement with the model near
$\epsilon=0$.  The deviation from a power law distribution at the low
moment range is associated with the reduced resolution of the
observational network for small events.  In mean field theory the
universal scaling function $f(x)$ in Eq. (\ref{eq6}) is of the exact
form \cite{fisher2}:$f_{mf}(x)=Axe^{-Bx^2/2}$ with non-universal
constants $A=B=1$. In FIG. \ref{Mplot} we present a collapse of the
observational data of $\la dm_0(t|M_0)/dt \ra$ for four different values of
$M_0$ to obtain the corresponding function $f_{exp}(x)$ for
observations with $x=t/M_0^{1/2}$. The observational curves not only
collapse, and are therefore {\em universal}, the mean field exponent
$1/2$ in the scaling variable $x$ is in excellent agreement with
observations. We fit the functional form $f_{mf}(x)$ with $A = 4$ and
$B = 4.9$ to the collapse of the observed data; $f_{exp}(x)$ deviates
from the $f_{mf}(x)$ for small values of the ordinate.

In mean field theory, the function $g(x)$ of Eq. (\ref{eq7}) is of the
symmetric form: $g_{mf}(x)=Ax(1-x)$, where $A$ is a non-universal
constant. In FIG. \ref{Tplot} we collapse observational data for
$\la dm_0(t|T)/dt \ra$ for three values of $T$ to obtain the function
$g_{exp}(x)$ with $x=t/T$.  Again we find that the curve collapses
well, even though only small data sets were available, and the exponent
of $1$ obtained from the collapse is in excellent agreement with mean
field theory.

We also plot the mean field scaling function $g_{mf}(x)$ with $A =
80$. The results show that while the scaling exponents agree, there
are notable differences between the observational function
$g_{exp}(x)$ and the mean field function $g_{mf}(x)$, especially for
small values of the ordinate. We checked that finite size effects do not
play a role in $g_{mf}(x)$ (or in $f_{mf}(x)$ for that matter).  Also,
we find that the mean skewness coefficient for the $g_{exp}(x)$
curves is 0.878 and the mean standard error of skewness is 0.705:
since twice the standard error is greater than the absolute value of
the skewness, the asymmetry is not statistically significant.
Therefore more observational work with a larger data set is required
to verify the moment rate shape asymmetry and clarify its origin.  An
asymmetry may result from a rupture process that begins with a failure
of a large asperity, from finite fault size effects if the
rupture process slows down once the rupture has traversed the fault in
one direction, or from contributions of early aftershocks. 

\section{\label{conclude} Discussion} 

The employed earthquake model \cite{benzion2,benzion4} was shown in
the past to have a critical point at $\epsilon=0$ and additional
dynamic regimes for $\epsilon>0$ \cite{fisher,dahmen} compatible with
observed frequency-size statistics of earthquakes on individual fault
zones \cite{wes,stirling}. We have generalized the theoretical
analysis to include a strengthening regime $\epsilon < 0$ with
aftershocks, and derived universal scaling functions around the
critical point $\epsilon=0$. The results provide new tools for data
analysis that may be used to obtain an improved understanding of
earthquake dynamics.  The analysis indicates that near $\epsilon=0$
the model is in the same universality class as a recent model for
domain wall motion in magnets and we can match the phase diagram
regions $\epsilon <0$ and $\epsilon >0$ to those of the 
corresponding magnet
model \cite{Nature,Matt1,durin,fisher}. In other words, the two
systems are marked by the exact same universal scaling exponents,
universal scaling functions, and similar phase diagrams. The model
predictions for frequency-size statistics and moment-rates of
earthquakes near $\epsilon=0$ are overall in close agreement with
observational data of relatively large earthquakes recorded on the global
seismic network \cite{bilek}.  However, the observed mean moment-rate of
earthquakes with a given duration $T$ apparently increases with time
more rapidly than it drops off, contrary to the corresponding
symmetric model function. The potential asymmetry in the observed data
may result from a rupture process that begins with a failure of a
large asperity. This is compatible with observations that hypocenter
locations tend to be located close to an area on a fault that produces
large moment release, (e.g. \cite{mai}). 
As we explain below, an analogy to magnetic systems suggests that
an asymmetry could potentially also stem from momentary initial 
threshold strengthening
associated with the creation of a new population of asperity contacts
upon local failure and followup aftershocks. 
Further study will be necessary to clarify this issue.

\begin{figure}
\includegraphics*[scale= .75]{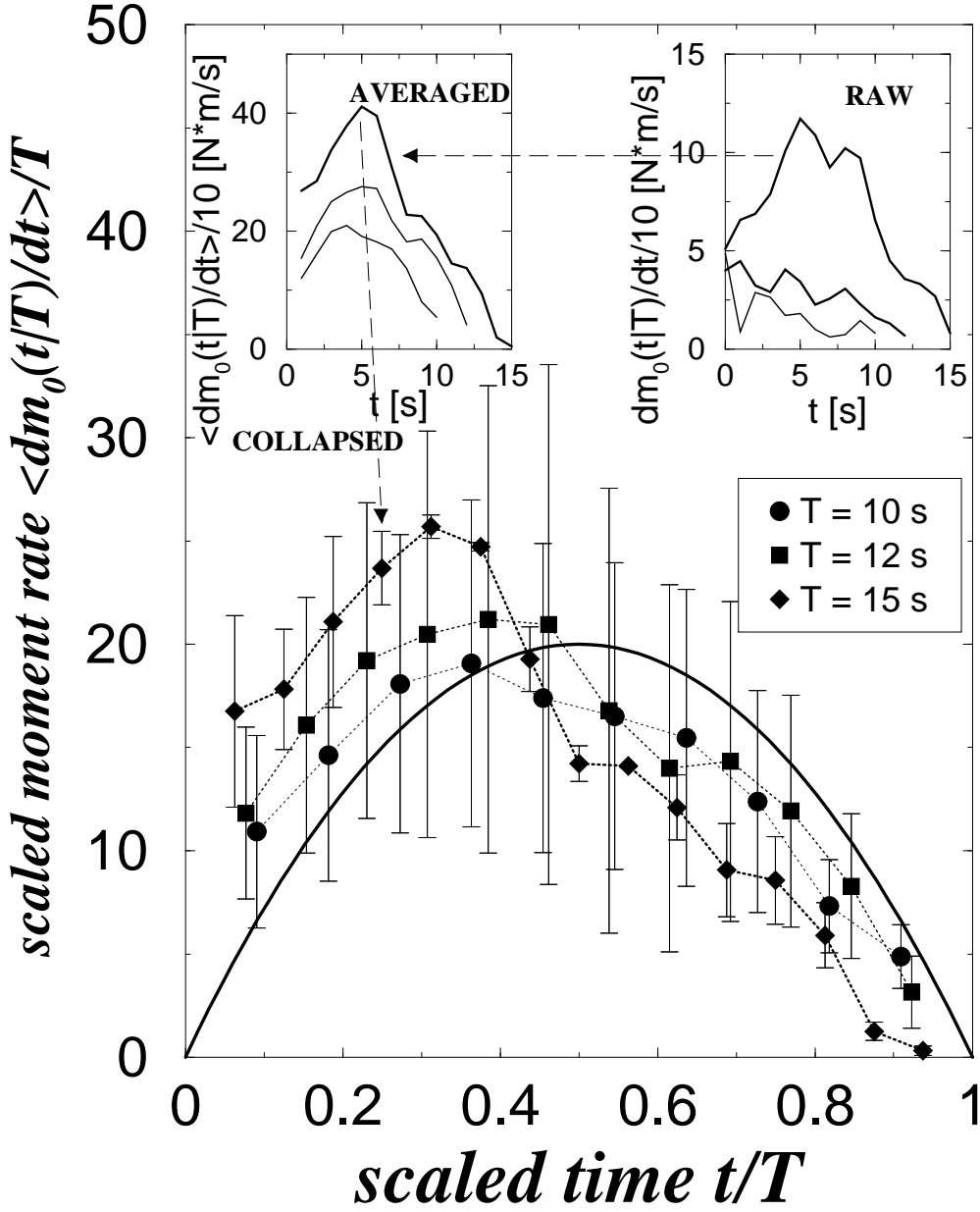}
\caption{\label{Tplot} A collapse of averaged earthquake pulse
  shapes, $\la dm_0(t|M_0)/dt\ra$ with a duration of $T$ (seconds)
  within 10$\%$ (given in legend), is shown. The collapse was obtained
  using the mean field scaling relation \protect\cite{Matt1}: $\la
  dm_0(t|T)/dt \ra\sim g(t/T)$ . In order to obtain each collapsed
  pulse shape, two to ten earthquakes were averaged for each value of
  $T$. In our mean field theory the universal scaling function is
  $g_{mf}(x)=Ax(1-x)$ with $x=t/T$. We plot this functional form (bold
  curve) with $A = 80$.  Note the apparent asymmetry to the left in
  the observed data while the theoretical curve is symmetric around its
  maximum. Inset: The raw data and the averaged data (before
  collapsed). }
\end{figure}

Theoretical analysis of such a potential asymmetric rupture process
requires corrections to the mean field earthquake model results. Recently
it has been shown that the corresponding magnetic domain-wall model
\cite{Nature,mehta} predicts well the critical scaling exponents for
Barkhausen noise experiments in magnets. Significantly, the
experimental scaling function for magnetization avalanches or
Barkhausen "pulses" \cite{Nature,mehta}, that is the analogue of the
moment-rate time profile for fixed earthquake duration of Eq.
(\ref{eq7}), shows the same type of asymmetry that is apparently
observed for earthquakes (FIG.  \ref{Tplot}). It has been
suggested that the asymmetry in the function is due to eddy currents
in the magnet \cite{Durin}.  Eddy currents have a similar effect in magnets as
transient threshold strengthening would have on earthquakes. In this
paper we have shown how long-term threshold strengthening (on time
scales longer than individual earthquakes) leads to aftershocks, which
effectively represent an asymmetry on time scales longer than
individual earthquakes. This raises the possibility that the origin of
this asymmetry may be similar in both magnets and earthquakes: in both
cases it may be
due to a transient force (due to eddy currents or threshold strengthening
respectively) that counteracts the propagation of an
event and thus leads to asymmetric event profiles that taper off more slowly
than they began.  Our
study shows that there are important theoretical and observational
connections between processes in earthquake and magnetic systems.

A related study of earthquake moment rate shapes was done by Houston
\cite{Houston} who used a data set similar to the 
data of Bilek \cite{bilek} used here. In order to
compare the average moment rate profiles of these earthquakes, 
the profiles were rescaled by
two methods: moment scaling and duration scaling. Moment scaling
rescales the time axis of the profiles using assumptions about
cracklike scaling of the events (different from our mean-field 
result), and then rescales the height
of the profiles so that the area (or moment) under all profiles
are the same. The duration scaling of \cite{Houston} rescales the
time axis such that all duration-scaled profiles end at the same 
reference duration, and then rescales the moment rates so 
that all scaled time profiles have the same area underneath. 

\begin{figure}
\includegraphics*[scale= .75]{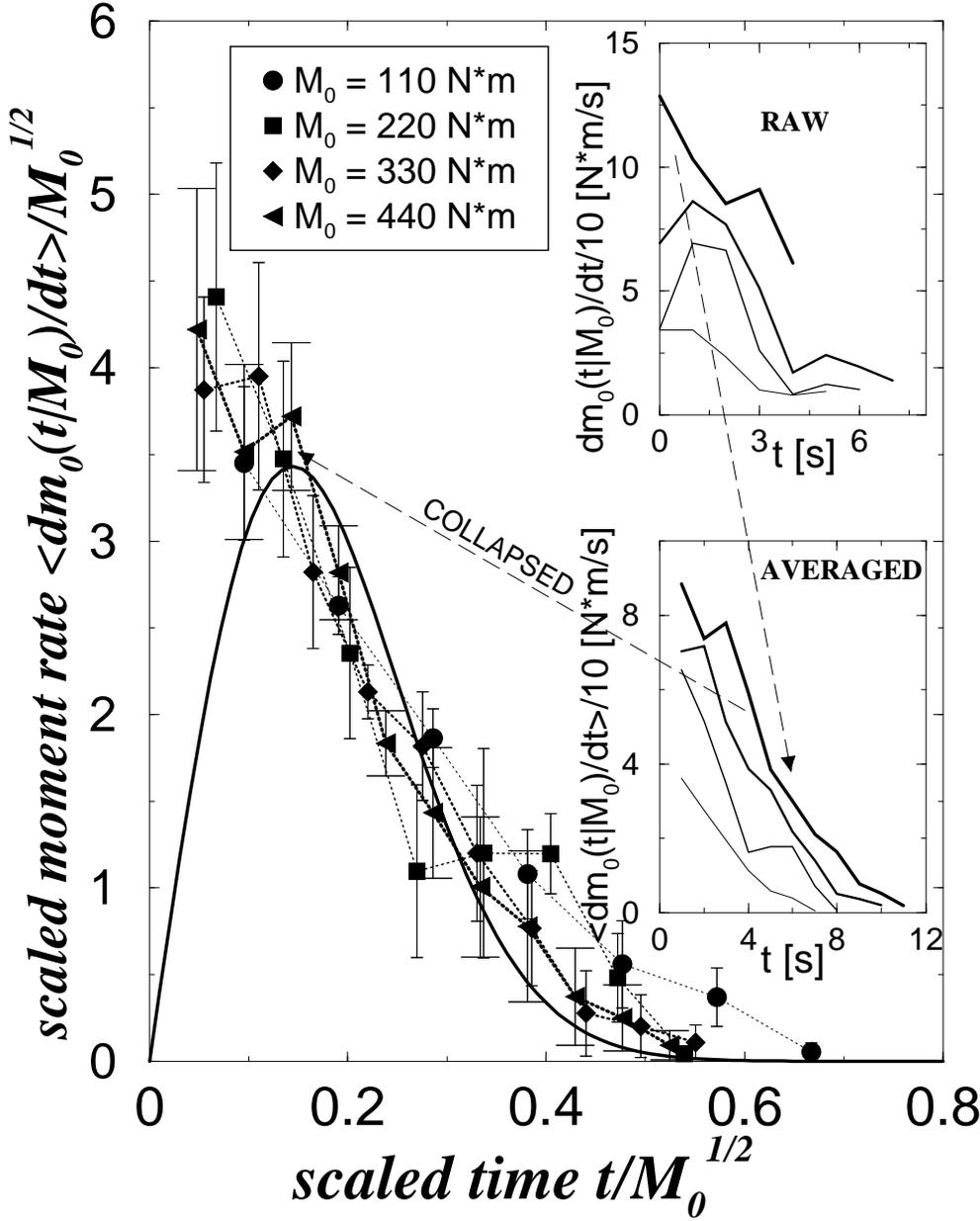}
\caption{\label{Mplot} A collapse of averaged earthquake pulse shapes,
$\la dm_0(t|M_0)/dt \ra$, with the size of the moment $M_0$ in Newton meters
within 10$\%$ of each size given in the legend respectively. In order
to obtain each collapsed moment rate shape, five to ten earthquakes
were averaged for each value of $M_0$.  The collapse was obtained
using the mean field scaling relation \protect\cite{fisher}:
$\la dm_0(t|M_0)/dt \ra/M_0^{1/2}\sim f(t/M_0^{1/2})$ (see text
Eq. (\protect\ref{eq6})). In our mean field theory the universal
scaling function is $f_{mf}(x)=Axe^{-Bx^2/2}$ where
$x=t/M_0^{1/2}$. We plot this functional form (bold curve) with $A=4$
and $B=4.9$. Inset: The raw data and the averaged data (before
collapsed).}
\end{figure}

\begin{figure}
\includegraphics[3,3][240,300]{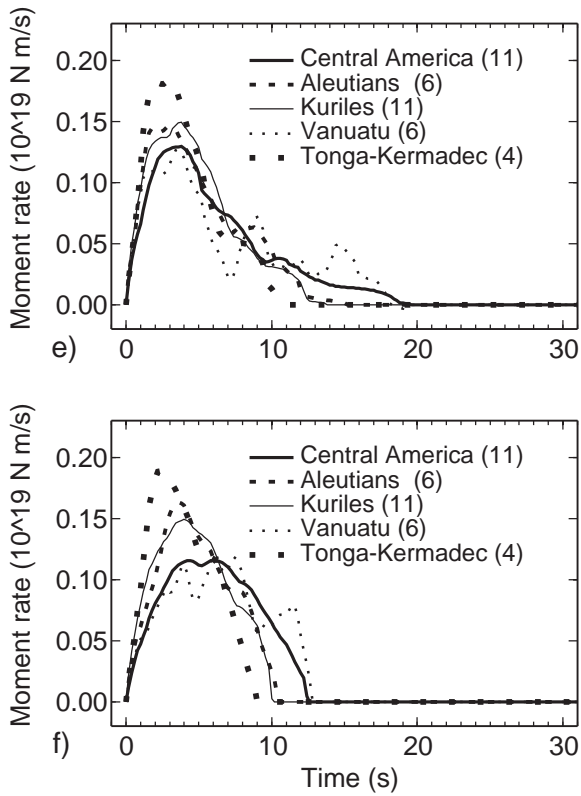}
\caption{\label{houstonfig} 
  Rescaled moment rate versus time profiles of seismic
  events determined with moment scaling (top) and duration scaling
  (bottom) obtained from \protect\cite{Houston} (see text). The data
  shown are obtained from several subduction zones in the indicated 
  geographical locations and  the numbers of earthquakes used for 
  each region are given in parentheses.}
\end{figure}

Fig. \ref{houstonfig} shows average moment rate profiles that were
obtained by Houston from moment scaling (top) and duration scaling (bottom)
for data from several subduction zones. 
The top plots correspond to $\la dm_0(t|M_0)/dt \ra$ collapsed, and the
bottom plots correspond to $\la dm_0(t|T)/dt \ra$ collapsed. We see from the 
top part in Fig. \ref{houstonfig} that overall the moment rate shapes seem to 
agree rather well with our predicted mean field shapes of Fig. \ref{Mplot}.
Likewise the bottom plots 
of Fig. \ref{houstonfig} appear to agree quite well the mean field 
curve of Fig. \ref{Tplot}, except for an additional slight asymmetry,
with positive skewness, similar to our result from observations.

In \cite{Houston} the skewness of the duration scaled moment rate profiles,
i.e. the skewness of the $\la dm_0(t|T)/dt \ra$  collapses, is calculated and
found to range from 0.12 to 0.36 depending on the depth of the
earthquakes (bigger skewness for greater depth) and the method used to extract
moment rates from seismograms. Since
the statistical error of the skewness is not given in \cite{Houston},
we cannot determine if the above skewness values are statistically
significant. It is interesting, however, that just like in the case
of magnets, and in our analysis of the observational data,
the skewness is slightly positive, indicating that on average earthquakes
tend to grow faster than they die down.
The skewness values quoted in \cite{Houston}
are smaller than the 0.878 value obtained here, though one of them 
(the 0.36 value for earthquakes at greater depth) is within errorbars.

While overall the shapes of the moment rate profiles look rather similar for 
both Houston's analysis and our mean field theory results, there are 
significant differences in the analysis that can be summarized as follows:

(1) The exponents used in \cite{Houston} to create the moment-scaled average
profile $\la dm_0(t|M_0)/dt \ra$ differ from the exponents we used. 
In both cases the $x$ axis is divided by $M_0^{\alpha_t}$ and the 
$y$ axis is divided by $M_0^{1-\alpha_t}$ in order to ensure normalization
of the curves to a fixed reference moment. In \cite{Houston} the exponent
value is $\alpha_t = 1/3$, while in this paper we use the mean field
value $\alpha_t=1/2$ (see Fig. \ref{Mplot}).
The value $\alpha_t = 1/3$ in \cite{Houston} is associated with cracklike
scaling; for cracks the moment $M_0$ scales with the 
rupture area $A$ as $M_0 \sim A^{3/2}$ \cite{benzion}. 
Since the rupture area scales
with the diameter $L$ of the crack as $A \sim L^2$ and the
duration $T$ of the earthquake scales with its diameter as 
$T \sim L$ one arrives at the proposed scaling $T \sim M_0^{1/3}$.
In \cite{Houston} it is mentioned, however, that the data would 
also be compatible with 
rescaling the time (or $x$) axis with an exponent of $\alpha_t \simeq 0.41$,
i.e. that $T \sim M_0^{0.41}$ for the data. This scaling   
is much closer to the mean field prediction $T \sim M_0^{1/2}$,
i.e. $\alpha_t = 1/2$, that is used in this work.

Our exponent $\alpha_t=1/2$ is derived from
the mean field prediction \cite{fisher} 
that the duration $T$ of earthquakes in the critical power law 
region of Fig. \ref{phase},
scale as $T \sim M_0^{1/2}$. In \cite{fisher} it is shown analytically that
the moment $M_0$ of earthquakes in the critical regime scales with 
the rupture area as $M_0 \sim A$. Since the rupture
area scales, as before, with the earthquake diameter as $ A \sim L^2$, and
the earthquake duration scales as $T\sim L$, one obtains for
the mean field prediction $T \sim M_0^{1/2}$ for critical earthquakes.
Note that for earthquakes which are so large that their horizontal 
diameter $L$ is much larger than the vertical width of the fault, 
one expects the moment to scale as $M_0 \sim L$.
Since again the duration scales as $T \sim L$, one obtains
$T \sim M_0$, i.e. $\alpha_t = 1$. The data used here and in \cite{Houston}
apparently does not fall within this scaling regime.

(2) Our theoretical analysis leads to predictions of both scaling exponents
and entire scaling functions, which can be compared to data.
In contrast, in \cite{Houston} the shape is obtained only empirically
by averaging rescaled data, and it is dependent on the chosen scaling 
exponents. Our procedure of comparing the theoretical results to data is 
designed to provide simultaneously, through a scaling collapse that minimizes
deviations of individual averaged curves from the collapsed one,
the scaling exponent $\alpha_t$ and the entire 
scaling function. On the other hand, in \cite{Houston} 
the value of $\alpha_t$ is
obtained from a separate analysis of duration versus moment of observed 
data, and the shape is then obtained empirically, by averaging rescaled 
data.

(3) For our simulated moment rates, the exponents used in \cite{Houston}
would not lead to a good collapse, {\it i.e.} the deviation between 
various rescaled curves would
be larger for $\alpha_t=1/3$ compared to $\alpha_t=1/2$.

Unfortunately, since the available observational data have large
error bars, we cannot determine at present unequivocally,
which set of exponents
works better. More moment rate data, especially for many small earthquakes
would allow us to reduce statistical errorbars and 
obtain a more precise
comparison of the scaling functions in theory and experiment.
The smaller earthquakes are also more likely to fall into the
critical scaling regime of our mean field prediction. However,
moment rate data are more difficult to obtain for small earthquakes,
as they require greater spacial resolution. 
We hope that this study will motivate more
observational work and analysis to answer these questions.

\begin{appendix}

\section{\label{Omori} Determination of Omori Law of Aftershock Decay}

We derive the Omori law for aftershock decay from our mean field
earthquake model in the case of dynamic strengthening.  After a
primary earthquake, the static failure stress, $\tau_s$, is increased
by an amount $\epsilon \la s \ra/N$, where $\la s \ra$ is the mean
event size. Given in terms of moment, the mean avalanche size in
mean field theory is simply given by \cite{dahmen}:
\be
\la s \ra = \int_{0}^{1/\epsilon^2} s/s^{3/2}ds \sim 1/\epsilon
\ee
where $1/s^{3/2} \sim D(s)$, is the scaling behavior of the mean field 
event size
distribution. Now since $\la s \ra \sim 1/\epsilon$, the failure
stress, $\tau_s$, is then shifted by an amount $1/N$.

In order to produce aftershocks, we assume that the new larger failure
stress, which we call $\tau_f^0$, is lowered slowly, logarithmically
in time until it returns to its original value, $\tau_s
\simeq \tau_f^0 - 1/N$.  More precisely, using  $t=0$ for the time at 
the end of the previous earthquake, $\tau_f^0$ decays as
follows:
\be \label{tauf}
\tau_f(t) = \tau_f^0(1-log(1+t))
\ee
We define $\eta(t)$ as the amount by which $\tau_f^0$ has been lowered
at time t:
\be \label{eta1}
\eta(t) = \tau_f^0-\tau_f(t) = \tau_f^0 log(1+t)
\ee
Eq. (\ref{tauf}) is only accurate to first order since lowering the
threshold triggers on average $\sim \eta(t)$ aftershocks (assuming
nonsingular stress distributions). Since aftershocks occur
quickly compared to the weakening time scale, the aftershocks that are
triggered due to lowering the threshold will cause $\tau_f(t)$ to
shift up by: $\eta(t) \times \epsilon \la s \ra/N \sim \eta(t)/N$.  
Thus the expression for $\tau_f(t)$ becomes:
\be
\tau_f(t) = \tau_f^0(1-log(1+t)) + \eta(t)/N
\ee
and Eq. (\ref{eta1}) gives:
\bea
&& \eta(t) = \tau_f^0 log(1+t)-\eta(t)/N\\
&& \eta(t) = \frac{\tau_f^0 log(1+t)}{1+1/N}
\eea
If we assume that the number of aftershocks triggered by time $t$ is
$A(t)\sim \eta(t)$, we obtain the  modified Omori law (Eq.(\ref{eq3}))
for aftershock decay
\cite{utsu,kiss,benzion}:
\be
\frac{\partial A(t)}{\partial t} \sim \frac{\tau_f^0}{(1+t)(1+1/N)}
\ee
For large $t$ and large $N$ we have the original Omori law:
\be
\frac{\partial A(t)}{\partial t} \sim \frac{\tau_f^0}{t}
\ee
\end{appendix}

\begin{acknowledgments}
We are grateful to Susan Bilek for giving us the
observational data.  We thank James P. Sethna and Michael B. Weissman
for very helpful discussions and MBW for first drawing our attention
to the possible similarity between the asymmetric universal scaling
function in magnets and earthquakes. A.M. and K.D. acknowledge support
from NSF grants No. DMR 03-25939 (Materials Computation Center), and
No. DMR 03-14279, the Alfred P. Sloan foundation (K.D.), and a generous
equipment award from IBM.  YBZ acknowledges support from a Mercator
fellowship of the German Research Society (DFG). YBZ and KD thank the
Kavli Institute of Theoretical Physics at UC Santa Barbara
for the hospitality during the final stages of this work and the partial
support through NSF under grant No. PHY99-07949.
\end{acknowledgments}

\break
\newpage

\end{document}